\documentstyle[12pt]{article}
\def\bib{\bibitem}
\def\be{\begin{equation}}
\def\ee{\end{equation}}
\def\beqar{\begin{eqnarray}}
\def\eeqar{\end{eqnarray}}
\def\barr{\begin{array}}
\def\earr{\end{array}}
\def\lsim{\:\raisebox{-0.5ex}{$\stackrel{\textstyle<}{\sim}$}\:}

\def\and{\qquad {\rm and } \qquad}

\def\p{\partial}

\def\slp{p \hspace{-1ex}/}
\def\sleps{ \epsilon \hspace{-1ex}/}
\def\slk{k \hspace{-1ex}/}

\def\prl#1{Phys. Rev. Lett. {\bf #1}}
\def\prd#1{Phys. Rev. {\bf D#1}}
\def\plb#1{Phys. Lett. {\bf B#1}}

\def\etal{ {\it et al.} }

\def\sbar{ \overline{s} }
\def\thmin{\theta_{\rm min}}
\def\cmin{\cos \theta_{\rm min}}
\begin{document}
\thispagestyle{empty}
\setcounter{page}{0}
\renewcommand{\thefootnote}{\fnsymbol{footnote}}

\begin{flushright}
MPI-PhT/94-21\\
PRL-TH/94-15 \\
April 1994
\end{flushright}

\vspace{5ex}
\begin{center}

{\Large \bf Test of $CP$ Violating Neutral Gauge Boson Vertices  in
$ e^+ e^- \rightarrow \gamma Z $ }\\

\bigskip
\bigskip
{\sc
   Debajyoti Choudhury$^{(a)}$\footnote{debchou@iws166.mppmu.mpg.de,
debchou@dmumpiwh.bitnet}
   and
   Saurabh D. Rindani$^{(b)}$\footnote{saurabh@prl.ernet.in}
   }

\bigskip
$^{(a)}${\it Max-Planck-Institut f\"ur Physik, Werner-Heisenberg-Institut,
F\"ohringer Ring 6, \\
80805 M\"unchen, Germany.} \\
$^{(b)}$ {\it Theory Group, Physical Research Laboratory, Ahmedabad--380 009,
India}

\bigskip
\bigskip
{\bf Abstract}
\end{center}

Forward-backward asymmetry in the process $e^+ e^- \rightarrow \gamma Z$
is proposed as a test for $CP$--violating anomalous $\gamma \gamma Z$ and
$ \gamma Z Z$ couplings.
Longitudinally polarized electron beams may be used to disentangle the effects
of these couplings.
We estimate possible limits that can be obtained from
$e^+ e^-$ colliders at centre-of-mass energies of 200 GeV and 500 GeV.

\newpage
\setcounter{footnote}{0}
\renewcommand{\thefootnote}{\arabic{footnote}}

The standard model (SM) has been found to be in extremely good agreement
with all experimental data to date. However, a direct precise
measurement of the form and magnitude of the gauge-boson self-couplings
has yet to be done, and has to wait for $e^+e^-$ experiments above the
$W^+W^-$ threshold proposed at LEP200 in the near future. While   there
has been considerable literature on measurement of possible
anomalous $\gamma W^+W^-$ and $ZW^+W^-$ couplings at both
hadronic \cite{hadronic} and leptonic colliders \cite{leptonic},
not much attention
has been paid to possible trilinear couplings among the neutral gauge
bosons $\gamma$ and $Z$, which are absent in SM to the lowest order in
the gauge coupling constants. These would be equally interesting to look
for at $e^+e^-$ colliders, possibly even at the existing ones. Examples
of processes where anomalous trilinear couplings of these neutral gauge
bosons would contribute are $e^+e^- \rightarrow \gamma \gamma ,\gamma Z,
ZZ$. These processes also have SM contributions to them through
$t$- and $u$-channel electron exchange diagrams. Experiments could
therefore compare the SM predictions with data and put bounds on the
anomalous couplings.

Of the processes mentioned above the process $e^+e^-\rightarrow \gamma
Z$ is all the more interesting for an additional reason. If $CP$ is conserved,
the
photon observed in the reaction should be produced symmetrically in the
forward and backward directions \cite{stodolsky}.
Thus the observation of a simple
forward-backward asymmetry of the photon emission direction would be a
signal of a $CP$-violating coupling. Such an asymmetry is far simpler to
observe accurately than more complicated energy and azimuthal
asymmetries of decay products
needed to observe $CP$ violation in the case of charged gauge bosons in the
final state \cite{biswarup}.

The above considerations have motivated us to look at tests of
$CP$-odd anomalous $\gamma\gamma Z$ and $\gamma ZZ$ couplings
in the reaction
\begin{equation}
e^-(p_-)+e^+(p_+)\rightarrow \gamma (k_1)+Z(k_2).
        \label{process}
\end{equation}
We calculate the forward-backward asymmetry of the photon (or the $Z$)
in the centre-of-mass (c.m.) frame in the presence of $CP$-violating
couplings, as well as the contribution of these couplings to the total
cross section. We then estimate possible limits on the couplings which can
come from future experiments at LEP200 and also at a future Next Linear
Collider (NLC) operating at a c.m. energy of 500 GeV\footnote{Forward-backward
asymmetry in this process has been considered in ref.\cite{czyz}. However, the
 interactions they consider do not obey electromagnetic gauge invariance and
Bose symmetry. Moreover, they were mainly concerned with results at LEP1 at
c.m. energy of 100 GeV. While they did point out the need for beam polarization
 in discriminating between various contributions to the asymmetry, they did
not discuss the results of polarization quantitatively, as we have done here.
}.
%

A general effective Lagrangian  $CP$-violating for $\gamma\gamma Z$
and $\gamma ZZ$
interactions, restricted to operators of dimension 6, can be written as
\be
\barr{rcl}
{\cal L} &= & \displaystyle
   e \frac{\lambda_1}{ 2 m_Z^2} F_{\mu\nu}
    \left( \p^\mu Z^\lambda \p_\lambda Z^\nu
          - \p^\nu Z^\lambda \p_\lambda Z^\mu
      \right)
       \\[2ex]
& & \displaystyle
      +\frac{e}{16 c_W s_W} \frac{\lambda_2}{m_Z^2}
       F_{\mu\nu}F^{\nu \lambda}
       \left(\p^\mu Z_\lambda + \p_\lambda Z^\mu   \right),
\earr
      \label{lagrangian}
\ee
where
$c_W=\cos \theta_W$ and $s_W=\sin \theta_W$, $\theta_W$ being the
weak mixing angle.

Equation (\ref{lagrangian}) represents
 the most general $CP$--violating Lagrangian
invariant under electromagnetic gauge transformations.
Terms involving divergences of the vector fields have been dropped from the
Lagrangian as they would not contribute when the corresponding particle is
on the mass shell, or is virtual, but coupled to a fermionic current which is
conserved. Since we will
 neglect the electron mass, the corresponding current can
be assumed to conserved.

The Lagrangian in eq.(2) must be understood in the sense of representing
effective interactions arising from some underlying fundamental theory,
and therefore, the couplings $\lambda_{1,2}$, or rather their momentum
space counterparts, would represent form factors, and are not really
constants. Furthermore, $\lambda_{1,2}$ could, in general, be complex
quantities. In SM,  of course, these $CP$-violating effective interactions
do not arise even at the one-loop level, and are therefore expected to be
extremely small.
%

The $CP$-violating couplings of the Lagrangian in eq.(\ref{lagrangian})
can induce both electric and weak dipole moments of fermions at the one-loop
level. A look at the chirality structure shows that,
to the leading order, only $\lambda_1$  contributes to the
the weak dipole moment.
On the other hand, the electric dipole moment receives contributions
from both $\lambda_2$ and $\lambda_1$, but is much
more sensitive to the former.
  With a suitable procedure \cite{burgess} for regulating the divergences
accompanying such a calculation with non-renormalizable interactions,
the contributions to the dipole moment are expected to be
\begin{equation}
d_f \sim \frac{\alpha}{16 \pi s^2_Wc^2_W}\frac{m_f}{m_Z^2}e\lambda\:
 \ln \frac{\Lambda^2}{m_Z^2}
   \label{dipolemom}
\end{equation}
where $\lambda$ is either $\lambda_1$ or $\lambda_2$ and $\Lambda$ is the
cutoff denoting the onset scale for new physics. Assuming
$\Lambda \sim 1\: {\rm TeV}$, eqn.(\ref{dipolemom}) gives for the
electron dipole moment a contribution of about
$\lambda \times  10^{-24}$ $e$ cm. This then translates to
a limit of about $| \lambda_2| \lsim 10^{-3}$ from the bound on the
electric dipole moment of the electron\cite{pdg}, known to be less than
about $10^{-27} e {\rm cm}$.
Similarly  $| \lambda_1|$ can be constrained from the weak dipole moment
\footnote{The authors of ref.\protect\cite{boudjema} have also considered
the contribution of one of the form factors to the electric dipole moment of
fermions. Our limits though are much stronger.}.
 However, these are
only indirect limits, and a correct estimate crucially depends on
assumptions about the structure of the theory.
Thus a  direct measurement of the couplings is
more desirable.

The Standard Model diagrams contributing to the process (\ref{process}) are
of course those with a $t$-- and a $u$--channel electron exchange, while
the extra piece in the Lagrangian (\ref{lagrangian}) introduces two
$s$--channel diagrams with $\gamma$-- and $Z$--exchange respectively.
 The corresponding matrix element is then given by
\begin{equation}
{\cal M}= {\cal M}_1 +{\cal M}_2 +{\cal M}_3 +{\cal M}_4 ,
       \label{amplitude}
\end{equation}
where
\be
\barr{rcl}
{\cal M}_1 & = & \displaystyle
   \frac{e^2}{4c_Ws_W}\, \bar{v}(p_+)\: \sleps(k_2) (g_V - g_A\gamma_5)
   \frac{1}{\slp_- - \slk_1}\: \sleps(k_1)\: u(p_-),
       \\[2ex]
{\cal M}_2 &=& \displaystyle \frac{e^2}{4c_W s_W}\, \bar{v}(p_+) \: \sleps(k_1)
  \frac{1}{\slp_- - \slk_2} \sleps(k_1) (g_V-g_A\gamma_5) u(p_-),
       \\[2ex]
{\cal M}_3 & = & \displaystyle \frac{ie^2\lambda_1}{4c_W s_W m_Z^2}\,
       \bar{v}(p_+)\gamma _\mu (g_V-g_A\gamma_5)u(p_-)
   \frac{(-g^{\mu\nu} + q^\mu q^\nu / m_Z^2)}
        {q^2-m_Z^2}
  V^{(1)}_{\alpha\nu\beta}(k_1,q,k_2) \epsilon^\alpha(k_1)
        \epsilon^\beta(k_2),
     \\[2ex]
{\cal M}_4 &= & \displaystyle
      \frac{ie^2\lambda_2}{4 c_W s_W m_Z^2}\,
   \bar{v}(p_+)\gamma _\mu u(p_-)
         \frac{(-g^{\mu\nu})}
              {q^2}
     V^{(2)}_{\alpha\nu\beta}(k_1,q,k_2)
      \epsilon^\alpha (k_1)\epsilon^\beta(k_2).
\earr
     \label{matrix elem}
\ee
We have used $q = k_1 + k_2$, and the tensors $V^{(1)}$ and $V^{(2)}$
corresponding to the three-vector vertices are given by
\be
\barr{rcl}
V^{(1)}_{\alpha\nu\beta}(k_1,q,k_2)
   &= &  \displaystyle
k_1\cdot q \: g_{\alpha\beta} \: k_{2\nu} + k_1\cdot k_2 g_{\alpha\nu} q_\beta
- k_{1\beta} \: q_\alpha \: k_{2\nu}  - k_{1\nu} \:q_\beta \: k_{2\alpha}
   \\[2ex]
V^{(2)}_{\alpha\nu\beta}(k_1,q,k_2)
   & = & \displaystyle
  \frac{1}{2}   \left[
  g_{\alpha\beta}
            \left( k_2\cdot q \: k_{1\nu} - k_1\cdot q \: k_{2\nu} \right)
        - g_{\nu\alpha}
            \left( k_2\cdot q \: k_{1\beta} + k_1 \cdot k_2 \: q_\beta \right)
      \right.
      \\[2ex]
   & &  \displaystyle \hspace{2em}
    \left.
       + g_{\nu\beta}
            \left( k_1\cdot k_2 \: q_\alpha - k_1\cdot q \: k_{2\alpha} \right)
        + q_\alpha \: k_{2\nu} \: k_{1\beta}
        + q_\beta \: k_{1\nu} \: k_{2\alpha}
    \right]
{}.
\earr
     \label{vertices}
\ee

In the above equations,
the vector and axial vector $Z$ couplings of the electron are given by
\begin{equation}
g_V = -1 + 4\sin^2\theta_W ;\quad g_A = -1
     \label{gVgA}.
\end{equation}

For compactness, we introduce the notation :
\be
\barr{rcl}
\sbar & \equiv & \displaystyle \frac{s}{m_Z^2}  \; ,
\\[2ex]
   {\cal B} & = & \displaystyle
      \frac{\pi\alpha^2}{8 s_W^2 m_W^2 \sbar}
     \left( 1 - \frac{1}{\sbar}   \right)
     (g_V^2+g_A^2) \; ,
\\[2ex]
C_{A} & = & \displaystyle
        \frac{\sbar - 1}{4}\:
    {\rm Im} \left( \lambda_1
                 - \lambda_2 \frac{g_V}{g_V^2+g_A^2} \right)  \; ,
\\[2ex]
C_{2L} & = & \displaystyle
                 \frac{(\sbar - 1)^2} {128} \:
       \left\{ |\lambda_1 |^2
               - 2 {\rm Re} (\lambda_1 \lambda_2^\ast)
                    \frac{g_V}{g_V^2+g_A^2}
               + \frac{ |\lambda_2 |^2} {g_V^2+g_A^2}
        \right\}  \; .
   \label{notation}
\earr
\ee

Using eqns. (\ref{amplitude} - \ref{notation}),
we obtain the differential cross section
for the process (1) to be
\be
\displaystyle
\frac{d\sigma}{d\cos\theta} =
    {\cal B}
\left[
       \frac{1}{\sin^2 \theta}
          \left( 1 + \cos^2 \theta + \frac{4 \sbar}{( \sbar - 1)^2}
           \right)
     + C_A \cos \theta
     + C_{2L} \left( \sbar + 2 + ( \sbar - 2) \cos^2 \theta    \right)
\right]  \; ,
    \label{diff c.s.}
\ee
where $\theta$ is the angle between photon and the $e^-$ directions.
The total cross section corresponding to the cut
$\thmin < \theta < \pi - \thmin$
can then be easily obtained by integrating the differential cross section
above and has been listed in Table 1 for two different values of
c.m. energy and
beam polarization (see later).

The presence of a term proportional to $\cos \theta$ in eqn. (\ref{diff c.s.})
reflects
the $CP$ violation in the interaction. It leads to a
forward-backward asymmetry given by the expression
\be
\barr{rcll}
A_{FB} & \equiv & \displaystyle
     \frac{F-B}{F+B}  &
    \\[2ex]
&= & \displaystyle
\frac{C_A}{2} \cos^2 \thmin & \displaystyle
  \left[ \left\{ \frac{\sbar^2 + 1}{(\sbar - 1)^2}
                   \ln \left( \frac{1 + \cmin }
                                   {1 - \cmin }
                            \right)
                 - \cmin
          \right\}
   \right.
\\[2ex]
& & & \displaystyle
\hspace{1.5em} + \left.
          C_{2L}
        \left\{ ( \sbar + 2) \cmin
               + \frac{ \sbar -2}{3} \cos^3\thmin
        \right\} \
    \right]^{-1} \; ,
\earr
\ee
where $F$ ($B$) represents the number of events with the photon in a
forward (backward) direction, with photon events lying within an angle
of $\theta_{min}$ from the beam axis excluded.

We have used $A_{FB}$ obtained above to estimate what limits can be put
on the imaginary parts of the anomalous couplings $\lambda_1$ and
$\lambda_2$ from the
future experiments at LEP200 with $\sqrt{s}=$200 GeV and at a possible
future linear collider with $\sqrt{s}=$500 GeV. For simplicity, we
neglect the real parts of $\lambda_{1,2}$, which appear in the total
cross sections, and henceforth $\lambda_{1,2}$ will refer to the
imaginary parts of the respective quantities. To get a
90\% confidence
limit on the anomalous couplings, we require that $F-B$ must exceed
$2.15\sqrt{N}$, where $N = F + B$ is the total number of $\gamma Z$ events
that can be observed with a certain integrated luminosity. We assume an
integrated luminosity of 500 pb$^{-1}$ for LEP200 and 10 fb$^{-1}$ for
the NLC. Our results are shown in Figs. 1 and 2, respectively for LEP200
and NLC. We have taken $\thmin$ to be $25^\circ$, as that gives the
largest sensitivity.

It is obvious from an examination of the expression for the
forward-backward asymmetry $A_{FB}$ that it is very insensitive to the
value of $\lambda_2$, since the latter comes multiplied by the
vector coupling
$g_V$ of the electron, which is very small ($\approx -.08$). This in turn
is a reflection of the fact that the $CP$ violating anomalous couplings
do not violate parity and therefore the interference of the photon
exchange term with the SM contributions can only contain $g_V$. This
insensitivity is clearly seen in the figures.

It is possible to enhance the relative contribution of the $\lambda_2$
term if the electron (and/or positron) beams can be longitudinally
polarized\footnote{It should be remembered, though, that polarization
of the electron beam alone  results in the initial state not
being an eigenstate of $CP$. However, since only opposite $e^+$ and $e^-$
helicities contribute in the limit of vanishing electron mass, the initial
state is effectively $CP$ even.
It can easily be seen that $O(\alpha )$ corrections with helicity flip
collinear photon emission, which survive even for vanishing electron mass,
do not contribute to $A_{FB}$.}.

This would give contributions which are proportional to
$g_A$, which is not small. It can be easily checked that the result of
including longitudinal polarizations $P_e$ and $P_{\bar{e}}$ of the
electron and positrons beams, respectively, is obtained from the
differential cross section of eq.(\ref{diff c.s.})
by the following replacements:
\be
\barr{rcl}
g_V^2 + g_A^2 & \rightarrow & g_V^2 + g_A^2 - 2 P g_V g_A,  \\
g_V & \rightarrow & g_V - P g_A,
\earr
     \label{polarization}
\ee
where $P = (P_e - P_{ \bar{e} } ) / (1 - P_e P_{ \bar{e} })$,
and multiplying the
differential cross section by an  overall factor of
$1 - P_e P_{ \bar{e} }$.

We have also shown in Figs. 1 and 2 the effect of electron polarization
$P_e =\pm 0.5$ assuming the positron to be unpolarized \footnote{ This
seems to us to be a conservative assumption as
the Stanford Linear Collider (SLC)
has already obtained a polarization of 62\%.
At least at linear colliders, it should certainly be possible to attain
a higher polarization.}. As expected, the slope of the bands change with
$P_e$. Experiments with polarized beams thus afford us complementary
studies and hence help unravel the effects due to the two couplings
in eqn. (\ref{lagrangian}). In case of a higher
polarization, the sensitivity of the experiments to $\lambda_2$ would be
much better.

It is seen that in general, limits of order 0.4--0.6 can be put on
individual couplings at LEP200, and of order 0.02--0.04 at NLC.
That the sensitivity increases with energy is easily understood from the
fact that the standard model contributions decrease with energy, whereas
the contribution of the anomalous couplings are constant with energy for
small values of the coupling, and eventually increase with energy.

Though our main aim was to
study the
$CP$-violating forward-backward
asymmetry, we have also looked at the  limits obtained on the
anomalous couplings from measuring the total number of events with a
cut of $25^\circ$  for the angle between the photon and beam directions. The
90\%
C.L. limits on the couplings are shown in Figs. 3 and 4, allowing for a
systematic uncertainty of 2\% in cross section determination. The
limits obtained are similar to those from the measurement of
forward-backward asymmetry\footnote{In fact, the limits derived from
total cross section are somewhat better for $\lambda_2$, but this feature
would disappear for higher beam polarization.}.
It should however be borne in mind that the
cross section could in principle receive contributions from several new
interactions, as for example, several $CP$ conserving $\gamma ZZ$ and
$\gamma \gamma Z$ couplings not included in our Lagrangian. Thus the
limits obtained are with the assumption that these couplings are absent.
The forward-backward asymmetry, on the other hand can receive
contributions only from the $CP$-violating interactions. Moreover, the
cross sections would also receive contributions from the real parts of
$\lambda_{1,2}$, which we have taken to be absent. Finally, systematic
uncertainties in normalization get cancelled out when one considers a
forward-backward asymmetry. Thus limits from the forward backward asymmetry
are more significant.

We should, of course, compare these limits from the bounds that can already
be inferred from limits on the radiative decays of the $Z$. The simplest
(naive)
possibility namely $Z \rightarrow \gamma \gamma$ of course does not exist
on account of Yang's theorem. Upper bounds on
$Br(Z \rightarrow f \bar{f} \gamma)$ however do exist \cite{pdg} and are
of the order of $5 \times 10^{-4}$.
The relevant matrix element is obtained from eqn(\ref{amplitude}) on
reversing the sign of the photon momentum, and the
extra contribution to the decay width given by
\be
\barr{rcl}\displaystyle
\delta \Gamma(Z \rightarrow f \bar{f} \gamma)
  &=  & \displaystyle
  \frac {\alpha^2 m_Z} {15360 \pi c_W^2 s_W^2} \:
               \left\{  (g_V^2+g_A^2 ) |\lambda_1 |^2
                  - 2 Q g_V {\rm Re} (\lambda_1 \lambda_2^\ast)
                  + Q^2 |\lambda_2 |^2
               \right\}
\\[2ex]
  &\approx  & \displaystyle
6.5 \times 10^{-7}\: {\rm GeV} \:
               \left\{  (g_V^2+g_A^2 ) |\lambda_1 |^2
                  - 2 Q g_V {\rm Re} (\lambda_1 \lambda_2^\ast)
                  + Q^2 |\lambda_2 |^2
               \right\}
    \label{decay width}
\earr
\ee
where $g_{V,A}$ are the corresponding couplings of $f$ to $Z$ as in
eqn(\ref{gVgA}) and $Q$ is its charge.
 (For hadronic final states, the right hand side in the expression above has
to be multiplied by 3.)\footnote{Note that $\delta \Gamma$ includes the
contribution
from the full available phase space unlike the experimental limits which have
kinematical cuts associated with them.}. As is easily seen, the limits imposed
by such decays are at best one (two) orders of magnitude weaker than the
bounds obtainable at LEP200 (NLC). In fact, the cleanest existing bound
(on $\lambda_1$) would come from the cross-section measurement for the
process $e^+ e^- \rightarrow \nu \bar{\nu} \gamma$ at LEP1 \cite{nunugamma}.
Translating the error bar in neutrino counting to a limit on on such couplings,
we have $\lambda_1 \lsim 4$.

Unitarity would constrain the high energy behaviour of the amplitude for the
process considered here. We find that the  constraint on the anomalous
couplings   obtained by demanding that unitarity not be violated until a scale
$\Lambda$ is weaker than that obtainable even from LEP200 so long as $\Lambda$
is not smaller than a few TeV.
%

It should be noted that in principle a $CP$-violating dipole type of
coupling of the electron to the photon or $Z$ could give rise to the
asymmetries we discuss. However, the corresponding contribution would be
proprortional to the electron mass, and would be negligible.

To summarize, we have pointed out an extremely simple $CP$-violating
effect that can be studied quite easily at forthcoming accelerators.
We have shown how anomalous trilinear neutral gauge boson couplings can
give rise to this effect. We have shown that limits of the order of
0.2--0.3 can be put on the imaginary parts of the corresponding form
factors at LEP200 with an integrated luminosity of 500 pb$^{-1}$ and of
the order of 0.02--0.03 at NLC with c.m. energy 500 GeV and integrated
luminosity of 10 fb$^{-1}$. Longitudinal polarization of the electron
beam can improve the sensitivity to the $\gamma\gamma Z$ coupling, which
is otherwise small.

{\bf Acknowledgements} We thank F.~Cuypers for sharing his plotting programmes.
SDR thanks the Max-Planck-Institut f\"ur Physik,
M\"unchen, for hospitality and the Indian
National Science Academy and Deutsche Forschungsgemeinschaft for financial
support.
\newpage
\noindent
{\large \bf Table Caption:}

\begin{enumerate}
\item
Standard Model $e^+ e^- \rightarrow \gamma Z$
cross-sections for various electron polarization $P_e$
(see eqn.\protect\ref{polarization}). The positron is
unpolarized.
\end{enumerate}

\vspace{2cm}

\noindent
{\large \bf Figure Captions:}

\begin{enumerate}
\item 90\% C.L. bounds on the $\lambda_1 - \lambda_2$ plane
($ \lambda_i \equiv {\rm Im} \lambda_i$) obtainable
from $A_{FB}$ at LEP200 with integrated luminosity of 500 pb$^{-1}$. The solid,
dashed and dot-dashed lines are for electron polarizations of 0, +0.5 and -0.5
respectively (positron is unpolarized).
\item As in Fig. 1 but for NLC ($\sqrt{s}$ = 500 GeV and
integrated luminosity of 10 fb$^{-1}$.
\item As in Fig 1, but for bounds from total cross-section alone.
\item As in Fig. 3 but for NLC.
\end{enumerate}

\newpage

\begin{table}[t]
$$
\begin{array}{|c||c|c|c||}
\hline
{\rm Energy} & \multicolumn{3}{c||}{ {\rm Cross Section (pb)} } \\
\cline{2-4}
& P_e = -0.5 & P_e = 0 & P_e = 0.5 \\
\hline \hline
\sqrt{s} = 200~{\rm GeV} & 8.49 & 7.93 & 7.38 \\
\hline
\sqrt{s} = 500~{\rm GeV} & 0.938 & 0.876 & 0.815 \\
\hline
\end{array}
$$
\caption{}
\end{table}

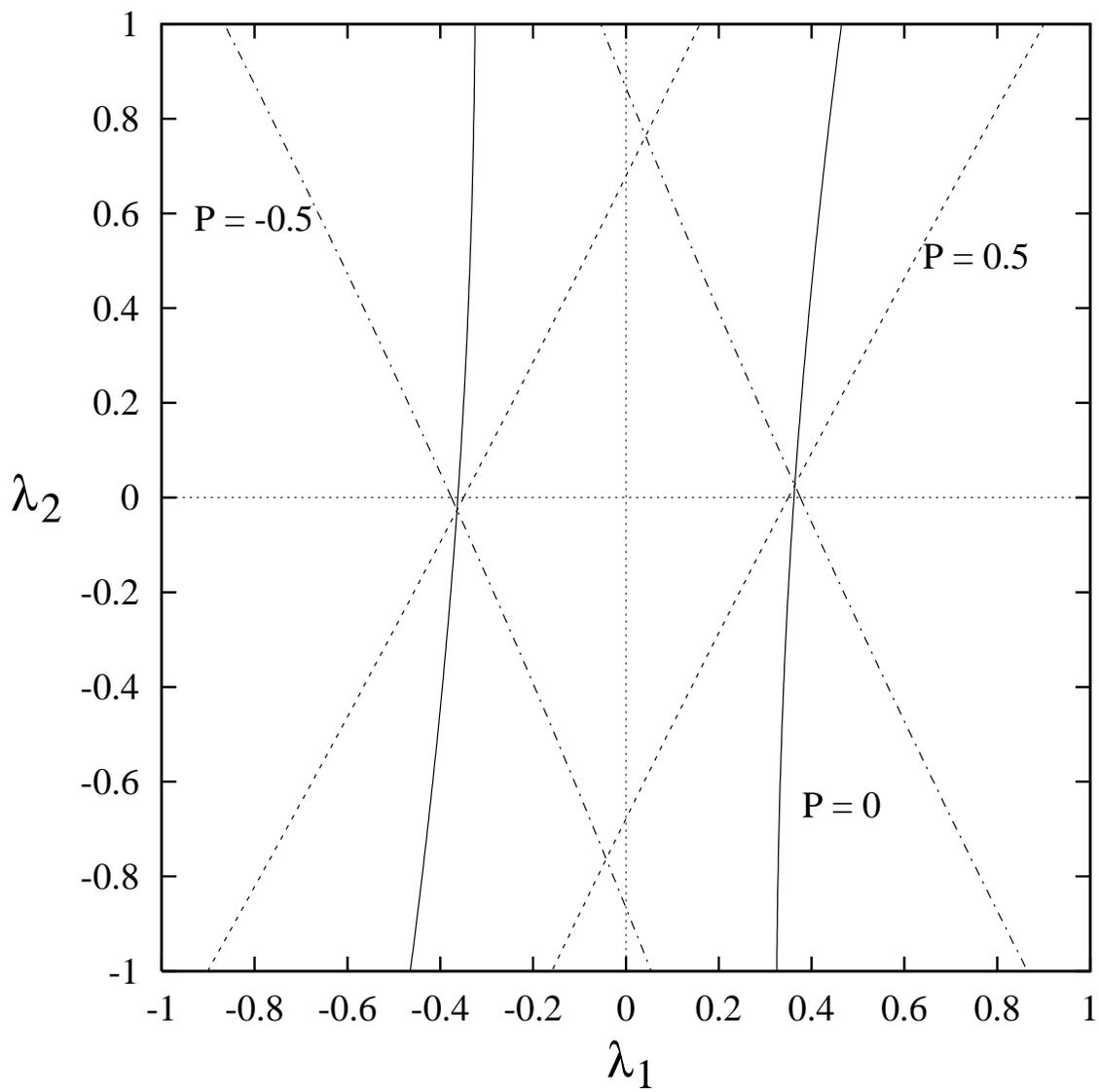
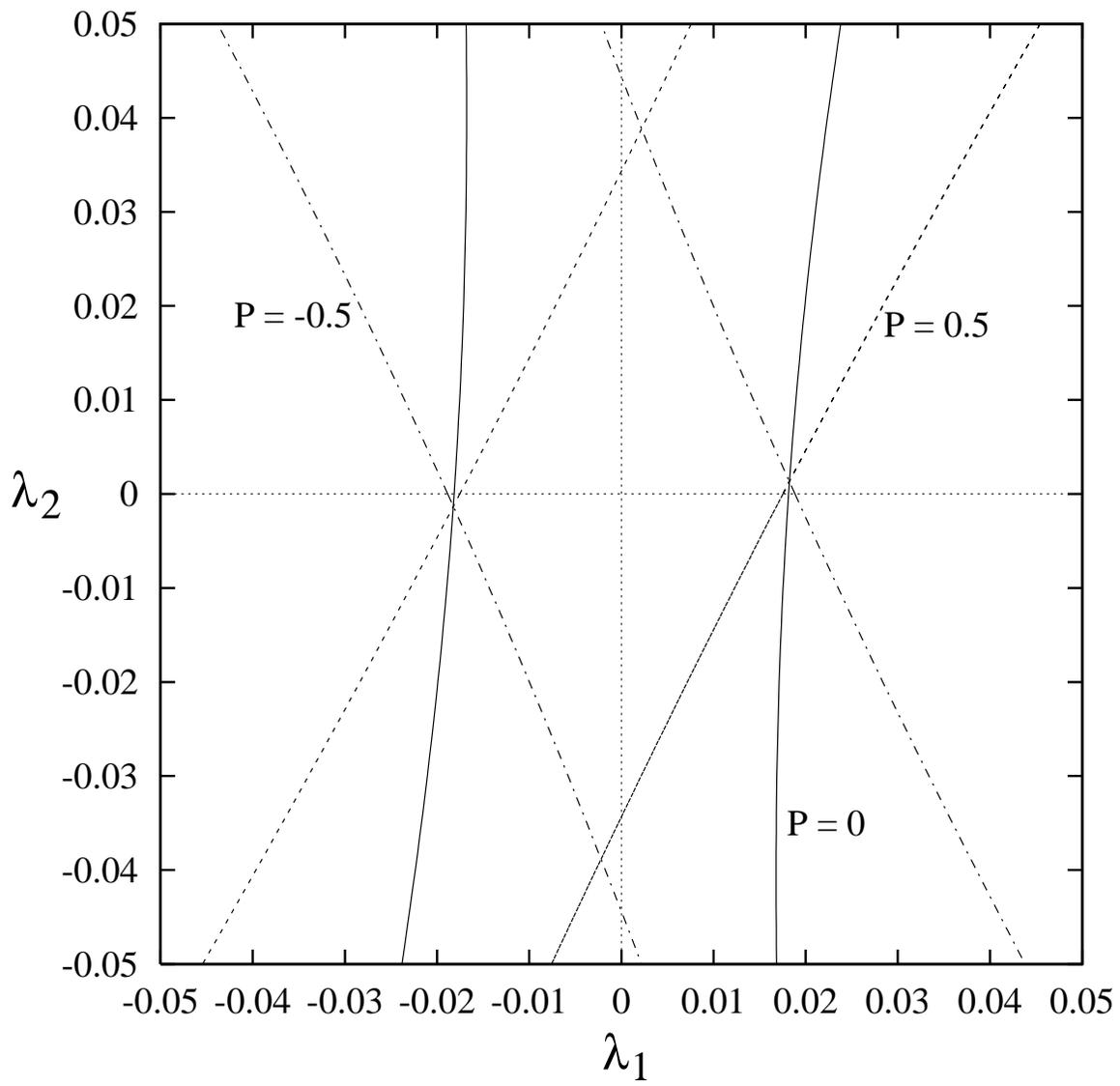
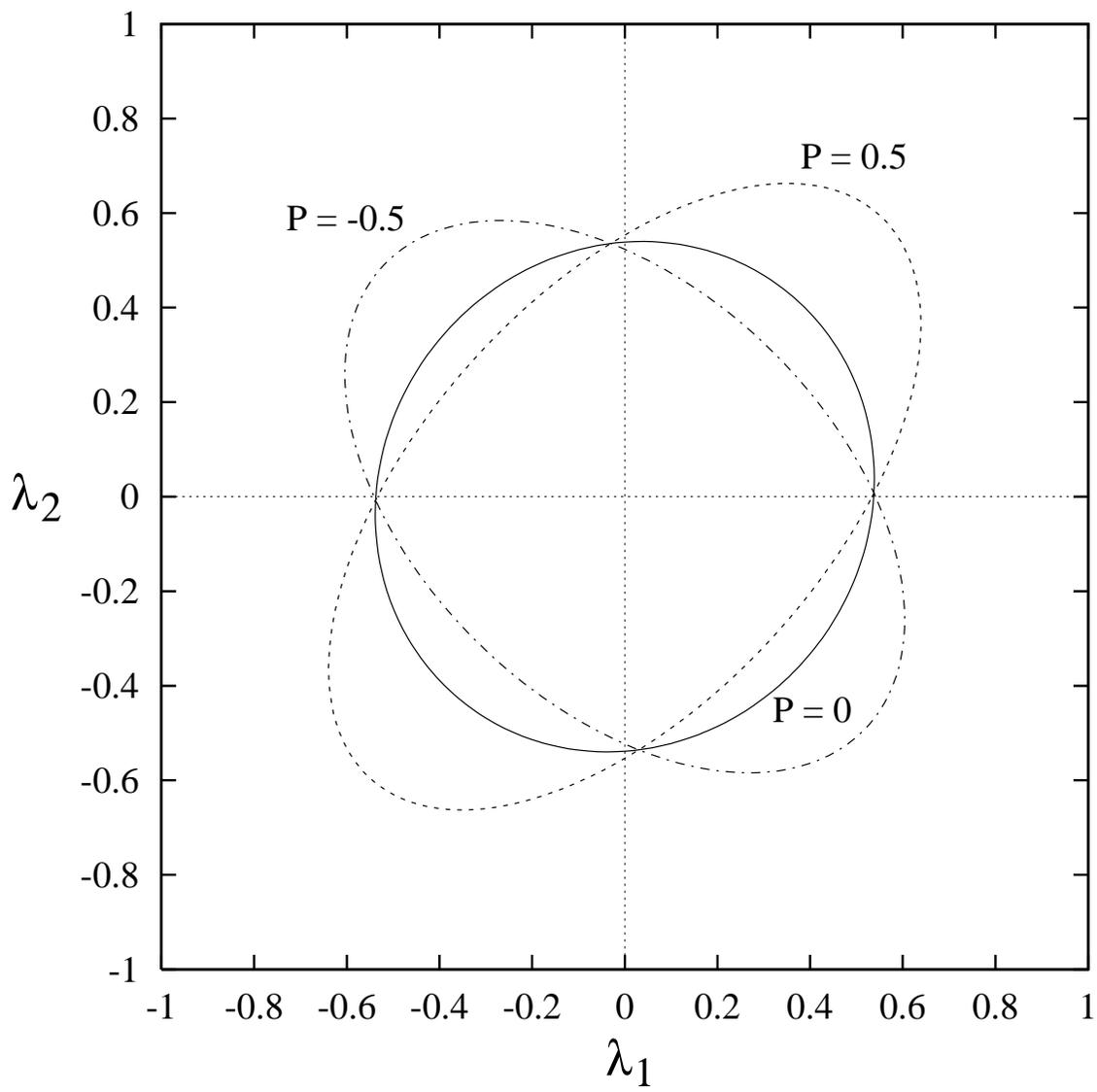
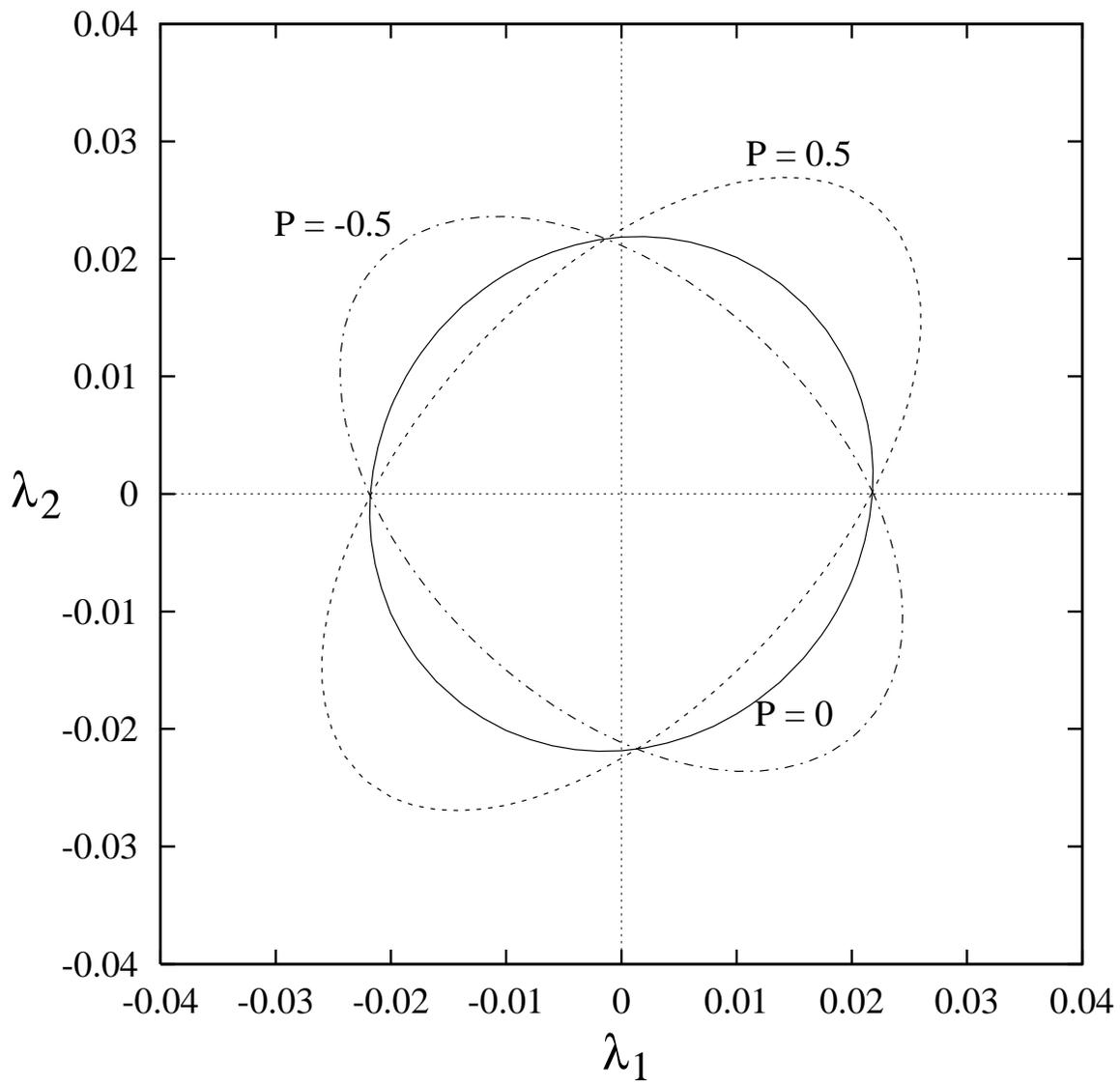
\end{document}